\documentclass[conference,10pt,twoside]{IEEEtran}

\usepackage{tikz}
\usetikzlibrary{arrows}
\usetikzlibrary{positioning}
\usepackage{mathtools,amsmath, amsthm, amssymb, amsfonts}
\usepackage[ruled,vlined]{algorithm2e}
\usepackage{chngcntr}
\DeclareSymbolFont{matha}{OML}{txmi}{m}{it}
\DeclareMathSymbol{\varv}{\mathord}{matha}{118}
\usepackage{amsmath,array,float}
\usepackage[export]{adjustbox}
\usepackage[utf8]{inputenc}
\usepackage{notoccite}
\usepackage{hyperref}
\usepackage[normalem]{ulem}
\usepackage{bm}
\usepackage[squaren, Gray, cdot]{SIunits}
\usepackage[bottom]{footmisc}
\usepackage[numbers]{natbib}
\usepackage{float}
\usepackage{ragged2e}       
\usepackage{booktabs,       
            makecell,       
            tabularx}       
\usepackage{cellspace}
\usepackage{balance}
\usepackage{relsize}%
\floatstyle{plaintop}
\restylefloat{table}

\usepackage{soul}
\usepackage{xcolor,graphicx}
\iftrue  
  \graphicspath{{img-color/}}
\else 
  \graphicspath{{img-gray/}}
\fi

\title{\Large \bf
Robust Learning-Based Sparse Recovery for Device Activity Detection in Grant-Free Random Access Cell-Free Massive MIMO: Enhancing Resilience to Impairments}

\author{\IEEEauthorblockN{Ali Elkeshawy\IEEEauthorrefmark{1}, Ha\"{\i}fa Far\`es\IEEEauthorrefmark{1}, Amor Nafkha\IEEEauthorrefmark{1} \\}
\IEEEauthorblockA{\IEEEauthorrefmark{1} IETR - UMR CNRS 6164, CentraleSup\' elec, avenue de la Boulaie - CS 47601 35576 \\ CESSON-SEVIGNE Cedex, France\\
Email: \{ali-fekry-ali-hassan.elkeshawy, haifa.fares, amor.nafkha\}@centralesupelec.fr}}

\DeclareUnicodeCharacter{2212}{-}
\begin{document}

\maketitle
\thispagestyle{empty}
\pagestyle{empty}
\begin{abstract}
    Massive MIMO is considered a key enabler to support massive machine-type communication (mMTC). While massive access schemes have been extensively analyzed for co-located massive MIMO arrays, this paper explores activity detection in grant-free random access for mMTC within the context of cell-free massive MIMO systems, employing distributed antenna arrays. This sparse support recovery of device activity status is performed by a finite cluster of access points (APs) from a large number of geographically distributed APs collaborating to serve a larger number of devices. Active devices transmit non-orthogonal pilot sequences to APs, which forward the received signals to a central processing unit (CPU) for collaborative activity detection. This paper proposes a simple and efficient data-driven algorithm tailored for device activity detection, implemented centrally at the CPU. Furthermore, the study assesses the algorithm's robustness to input perturbations and examines the effects of adopting fixed-point representation on its performance.
\end{abstract}

\maketitle
\IEEEdisplaynontitleabstractindextext
\IEEEpeerreviewmaketitle
\section{Introduction}
The rapid growth of massive machine-type communication (mMTC) has introduced significant challenges for next-generation wireless networks, particularly in accommodating an ever-increasing number of connected devices. Grant-free random access has emerged as a promising solution to address these challenges by allowing devices to transmit data without prior access requests, thereby reducing signaling overhead and latency \cite{Shahab2020}. However, this paradigm requires efficient and reliable device activity detection, especially in scenarios where non-orthogonal pilot sequences are employed, which further complicates the detection process.

From another hand, the increasing complexity driven by
the need to support this massive connectivity, has highlighted the limitations of traditional approaches in addressing challenges like sporadic activations, constrained resources, and interference. In this context, machine learning has emerged as a powerful tool, offering innovative solutions for activity detection and resource allocation \cite{bai2018deep} \cite{deSouza2023DeepLearning}.
This paper explores the application of supervised learning to
Grant-Free Random Access (GF-RA) in Cell-Free Massive
MIMO (CF-mMIMO) networks, showcasing its potential to
overcome these challenges through data-driven algorithms.
Unlike conventional methods such as maximum likelihood (ML) detection \cite{Ganesan2020Algorithm, ganesan2020clustering}, the proposed approach leverages machine learning capabilities to handle the inherent uncertainties of real-world wireless systems. Specifically, the detector demonstrates robustness against two major sources of impairments:\textbf{(i)}
imperfect channel knowledge due to the channel estimation errors. In realistic scenarios, it is almost impossible to provide a perfect channel estimator. \textbf{(ii)} CF-mMIMO systems are expected to make use of low cost components and low resolution analog-to-digital converters (ADCs) at both the access points and device equipments \cite{Cell-free-lowADC}. Consequently, to ensure optimal performance in detecting device activities, it's imperative to address the effects of low-bit quantization on the samples gathered at the APs. Simulation results demonstrate that the proposed deep learning-based activity detector exhibits superior performance compared to traditional maximum likelihood techniques in terms of accuracy and resilience. These qualities make it well-suited for providing reliable connectivity in real crowded mMTC scenarios.

The rest of the paper is organized as follows. Section II provides the system model. Section III describes the data-driven device activity detector. Section IV provides numerical results and a discussion. Finally, section V draws the conclusion of the paper.

\begin{figure}[t]
  \centering
  \includegraphics[width=\columnwidth, height=5cm]{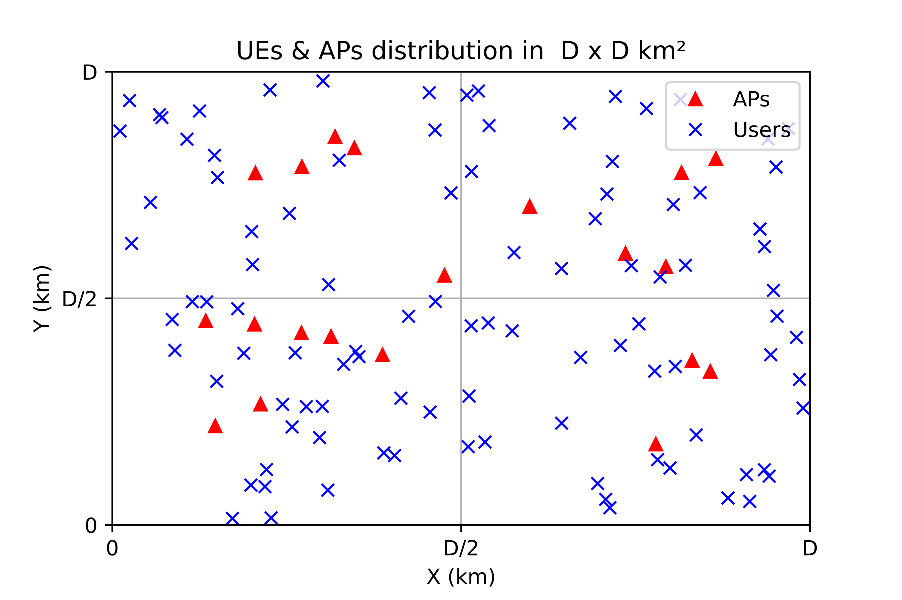}
  \vspace{-15pt} 
  \caption{Distribution of devices and access points within the network.}
  \label{Fig:UEs_APs_distribution}
\end{figure}

\begin{figure*}[t]
  \centering
  \includegraphics[width=\textwidth]{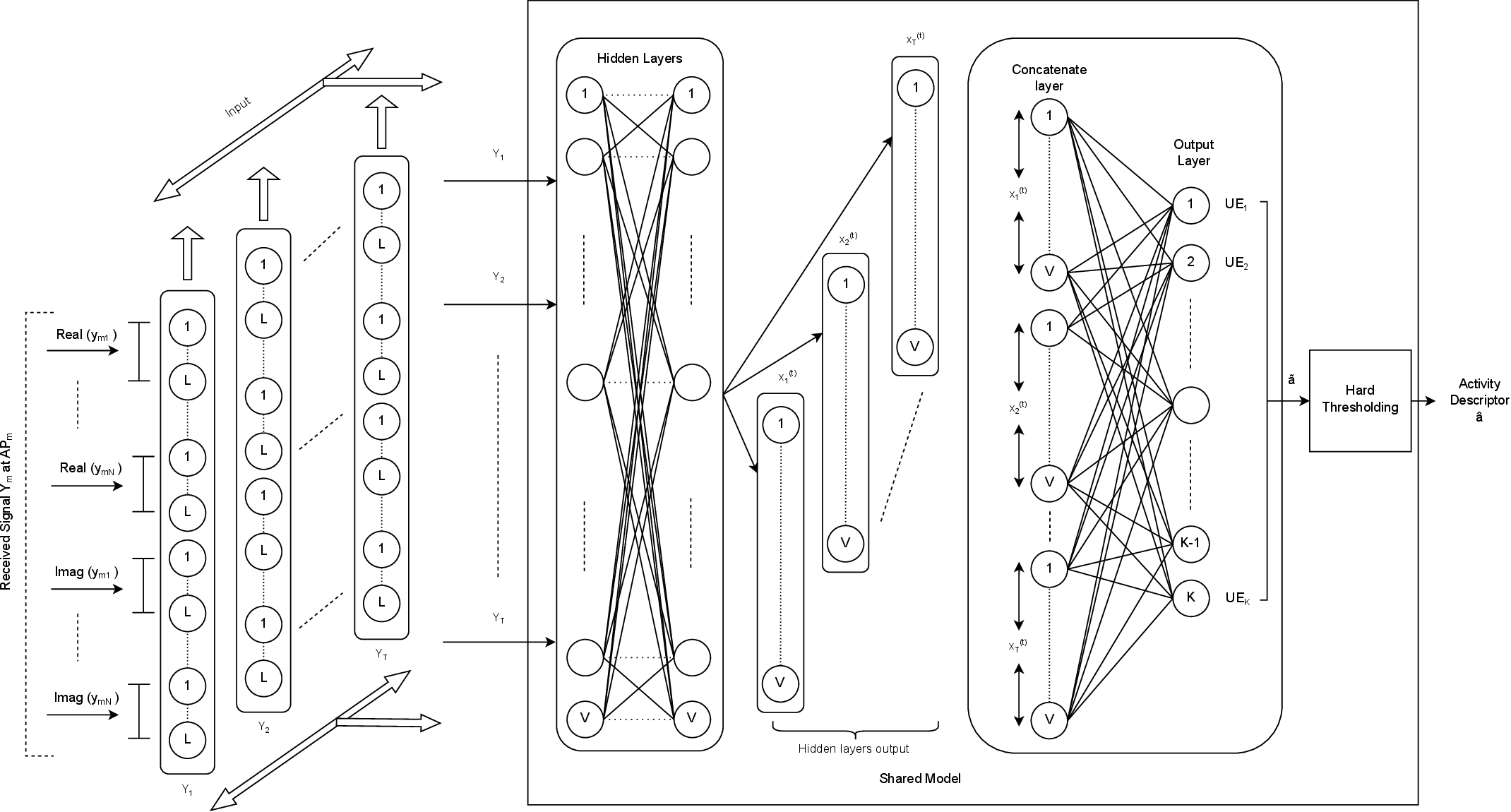}
  \caption{Architecture of the DMLP algorithm proposed for device activity detection.}
  \label{Fig:DLMP}
\end{figure*}

\section{System Model}
\label{Section:SystemModel}
We consider a CF-mMIMO network comprising $M$ APs, each equipped with $N$ antennas, serving $K$ single-antenna devices. Moreover, we assume that the APs and devices are independently and uniformly distributed within a square area of size $D \times D$ $\meter\squared$, as depicted in Fig. \ref{Fig:UEs_APs_distribution}, with the edges wrapped around to eliminate boundary effects. All the APs are connected to a central processing unit (CPU) through high-speed and infinite-capacity fronthaul connections. It is assumed that the devices are synchronized and simultaneously served by a cluster of APs of finite size, $T$, using the same time-frequency resources. Each device exhibits a probability, $\varepsilon \ll 1$, of transitioning to an active state. 

Let $a_k \in \{0, 1\}$, where $a_k = 1$  indicates that the $k$-th device is active, and $a_k = 0$ indicates it is inactive. Therefore, the device's activity status is defined by following probabilities: $\Pr(a_k = 1) = \varepsilon$ and $\Pr(a_k = 0) = 1 - \varepsilon$. Let $\mathbf{a} = (a_1, a_2, \dots, a_K)$ denotes the activity status of $K$ devices at any given time instant. Then, the received symbol vector $ \mathbf{y}_{mn} \in \mathbb{C}^{L \times 1}$ at the $n$-th antenna of the $m$-th AP can be expressed as:
\begin{equation}
\mathbf{y}_{mn} =  \sum_{k=1}^{K} \sqrt{\rho_k} a_k g_{mnk} \mathbf{s}_k + \mathbf{w}_{mn},
\label{eq:RxSig}
\end{equation}
where $\rho_k$ denotes the power transmitted by device $k$ when active, $\mathbf{s}_k \in \mathbb{C}^{L \times 1}$ represents the pilot sequence for the device $k$, and $\mathbf{w}_{mn} \in \mathbb{C}^{L \times 1}$ denotes an independent and identically distributed (i.i.d.) noise samples at the n-th antenna of the m-th AP, following the complex Gaussian distributions $\mathcal{CN}(0, \sigma^2\mathbf{I}_L)$ with a variance of $\sigma^2$ per dimension. The channel gain between the $k$-th device and the $n$-th antenna of the $m$-th AP is expressed as $g_{mnk} = \sqrt{\beta_{mk}} h_{mnk}$, where $\beta_{mk}$ represents the large-scale fading, accounting for both path loss and log-normal shadowing, and $h_{mnk} \sim \mathcal{CN}(0,1)$ denotes the small-scale fading. In this paper, we assume a block fading model, where each channel remains constant over a coherence interval, and all channels are independently distributed.

We assume that the large-scale fading coefficients $\{\beta_{mk}\}_{k=1\dots K}^{m=1\dots M}$ 
are known at the CPU \cite{Ganesan2020Algorithm,Beta1,Beta2}, which is not true in practical systems. Based on this information, each device $k$ is assigned a cluster comprising the best $T$ access points.
The large-scale fading coefficient $\beta_{mk}$ is the same as considered in \cite{ganesan2020clustering}, it follows the micro-cell propagation model given by:
\begin{equation}
\beta_{mk}[\text{dB}] = -30.5 - 36.7 \log_{10}\left(\frac{d_{mk}}{1\meter}\right) + F_{mk},
\end{equation}
where \( F_{mk} \sim \mathcal{N}(0, \sigma^2_{sh}) \) represents the shadow fading component with variance \( \sigma^2_{sh} \), and \( d_{mk} \) denotes the horizontal distance, in meters, between the \( k \)-th device and the \( m \)-th AP, neglecting their height differences.

In our case, assigning unique orthogonal pilot sequences to devices is impractical due to the limited duration of the channel coherence time, $\tau_C$. As a result, the system experiences pilot contamination, a challenge that occurs when active devices transmit non-orthogonal pilot sequences $\mathbf{s_k}$ during random access slots. These sequences, modeled as $\mathbf{s_k} \in \mathbb{C}^{L \times 1}$, are randomly selected from a Gaussian distribution. The length $L$ corresponds to a specified fraction of the coherence block, for instance, $\delta\%$.

\section{Learning-Based device Activity Detection}
\label{Section:DL_UAD}
In the proposed approach, the centralized activity detection model is executed on the CPU, independently of the network's access points. As illustrated in Fig..~\ref{Fig:DLMP}, its architecture includes a signal selection block designed to enhance processing efficiency. This architecture begins with an input layer that processes signals received from all APs within the selected cluster. It includes $Z$ fully connected hidden layers, each comprising $V$ neurons. A concatenation layer then combines and adjusts the $T$ outputs produced from previous hidden layers based on the processed signals. Finally, the output layer, containing $K$ neurons, generates the predicted activity vector. The hidden layers of the proposed model utilize the ReLU (Rectified Linear Unit) activation function, a popular choice for feedforward neural networks. The ReLU introduces non-linearity into the model, which is essential for capturing complex relationships in data, while maintaining computational efficiency. Additionally, its compatibility with backpropagation techniques effective training of deep and intricate network structures. The output layer employs the sigmoid activation function, which was specifically chosen for its effectiveness in binary classification scenarios. The aforementioned function provides a robust mechanism for discriminating between active and inactive device modes, thus proving highly suitable for accurate device activity status detection.

The detailed processing is as follows: the APs transmit their collected signals to the CPU, where the input comprises data from all $M$ APs in the network. The model processes this data in parallel for each device. Using large-scale fading coefficients, the selection block identifies the most relevant AP signals for each device, choosing the top $T$ signals $[\mathbf{Y}_1, \mathbf{Y}_2, \dots, \mathbf{Y}_T]$ for device $k$, as shown in Fig.~\ref{Fig:DLMP}. These selected signals are passed through the model's hidden layers, where $T$ feature vectors are extracted and combined in a concatenation layer. This process ensures that activity detection is specific to each device’s AP cluster, even though some clusters may overlap.

For each device $k$, the model isolates their activity while disregarding the activities of the other $K - 1$ devices, producing a predicted activity $\tilde{a}_k$. These predictions are aggregated to form an overall activity vector $\tilde{\mathbf{a}}$. For instance, the proposed model is designed to address a multi-label classification problem by minimizing a binary cross-entropy loss function, expressed as:
\begin{equation}
\mathcal{L}(\mathbf{a}, \hat{\mathbf{a}}) = - \sum_{k=1}^{K} \big[ a_k \log(\hat{a}_k) + (1 - a_k) \log(1 - \hat{a}_k) \big].
\end{equation}

The CPU then applies a hard decision rule with a threshold $\tau \geq 0$ to binarize the predicted activities, resulting in the final estimated activity vector $\hat{\mathbf{a}}$. The value of \(\tau\) is carefully chosen to achieve a specified false alarm rate, ensuring reliable activity detection.
This approach decouples the clustering process from the deep learning model, enhancing both flexibility and efficiency. Clustering is dynamically handled by the CPU before model predictions, reducing computational complexity while maintaining high detection accuracy for cell-free networks.

\begin{figure}[t]
  \centering
  \includegraphics[width=\columnwidth, height=6cm]{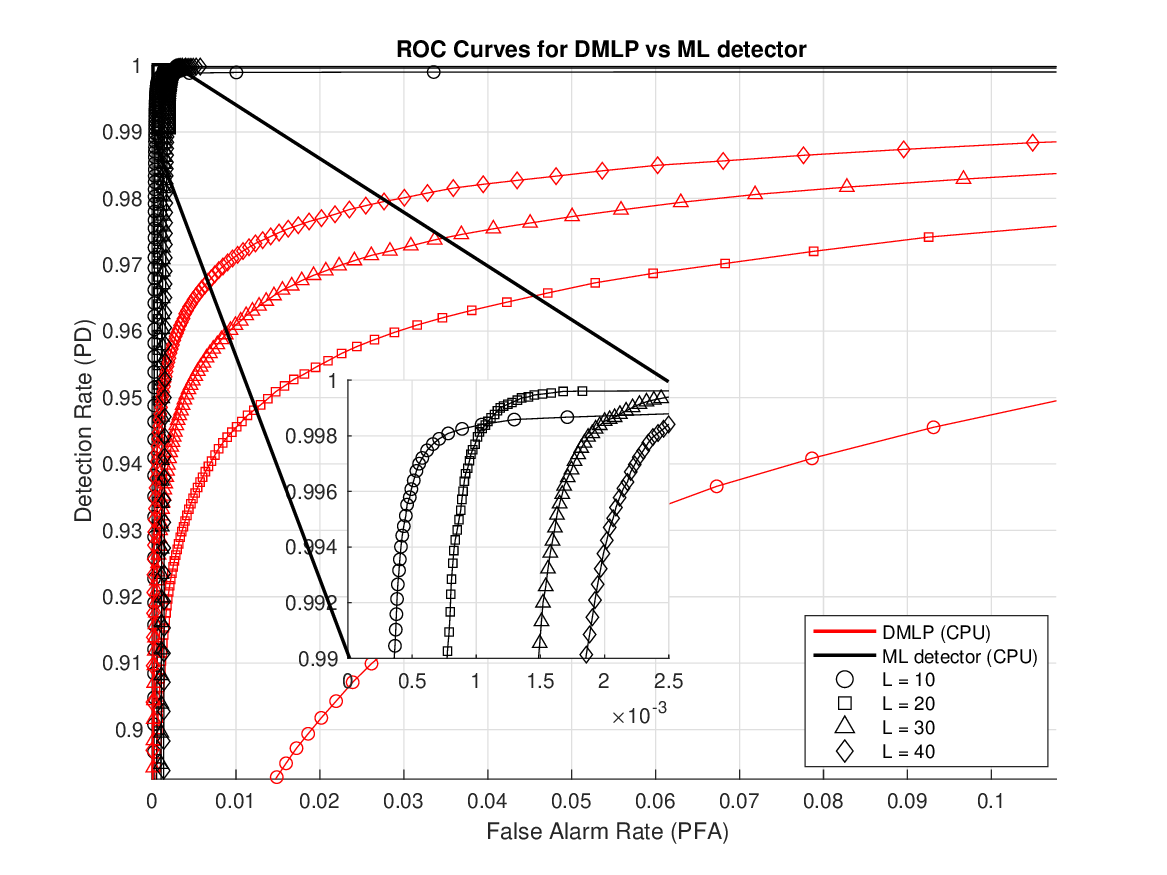}
  \vspace{-15pt} 
  \caption{ROC curve analysis for different pilot sequence lengths $L$.}
  \label{fig2}
\end{figure}

\begin{figure}[t]
  \centering
  \includegraphics[width=\columnwidth, height=6cm]{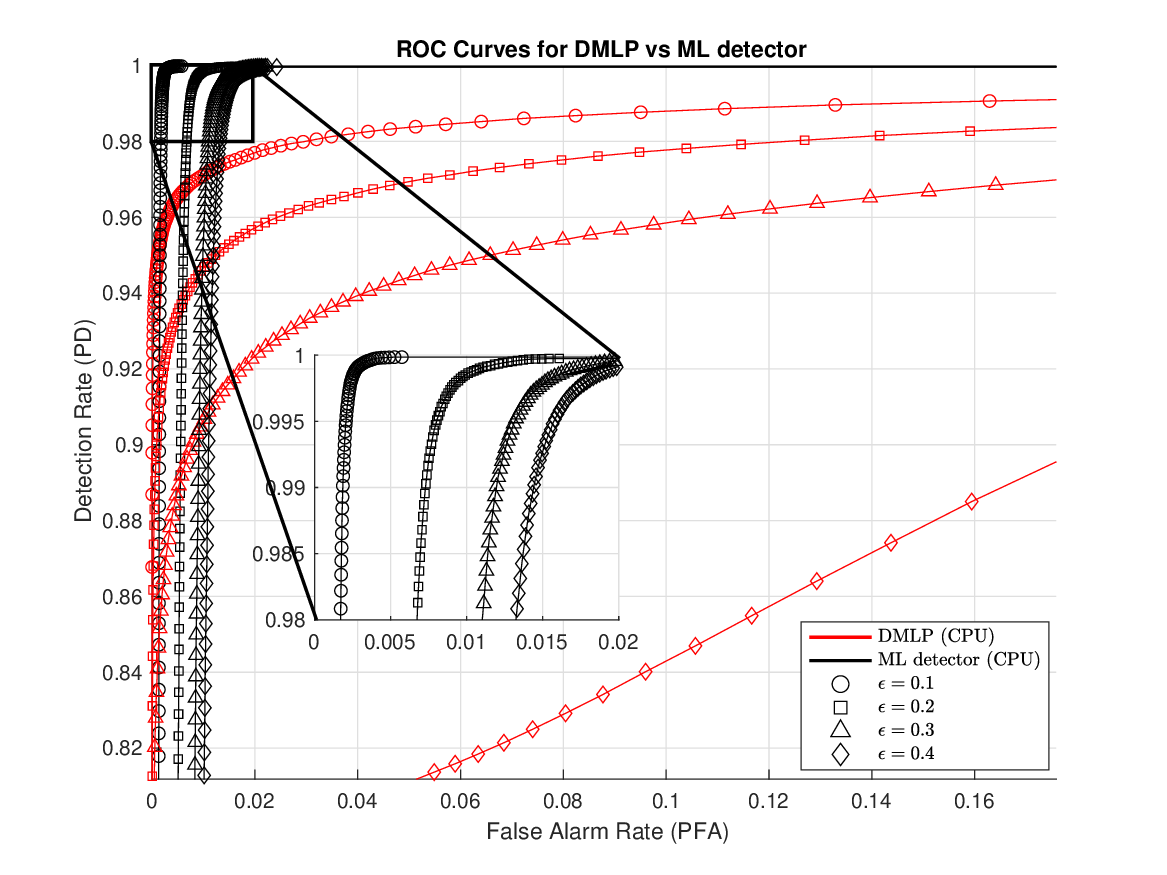}
  \vspace{-15pt} 
  \caption{ROC curve analysis for different sparsity degrees $\epsilon$.}
  \label{fig3}
\end{figure}

\section{Numerical Results}
\label{Section:NumResults}

\subsection{Simulation setup}
We analyze a square area configured with edge-wrapping to emulate an infinite network and eliminate boundary-related distortions. The area side is given by $D=1\kilo\meter$ and the edge distance is fixed to $50\meter$. We consider $M=20$ APs (each equipped with $N=2$ antennas) with a minimum spacing of $15\meter$, $K=100$ devices with a minimum spacing from APs of $10\meter$, and a cluster size $T=4$. 
In this simulation setup, we consider an activation probability $\epsilon = 0.1$, and a pilot sequence length $L = 40$. The propagation model adheres to a micro-cell framework, with the large-scale fading coefficient $\beta_{mk}$ derived as outlined in~\cite{Bjornson2019}. For small-scale fading, the Rayleigh model is considered.
Furthermore, the noise power is $\sigma^2=−109$ dBm, and we consider a coherence block of $1\milli \second$ and $200\kilo\hertz$, such that \(T_c=200\) symbols can be transmitted.

The upcoming analysis will rigorously compare the performance metrics of the proposed approach with those of the maximum likelihood activity detector (ML-detector) presented in \cite{ganesan2020clustering}. Finally, we employ $5\cdot10^4$ samples for the training dataset and $ 2\cdot10^4$ samples each for the validation and test datasets.

\subsection{Analysis without Uncertainty Considerations}
Before analyzing the performance metrics of the proposed approach, a hyperparameter tuning process was performed attain optimal precision in activity detection while concurrently minimizing the potential for overfitting. Specifically, we focus on deep architectures, where the number of hidden layers $Z$ was set to be greater than or equal to two, and $V$ indicates the number of neurons within each hidden layer. The optimal parameters were identified using the Pareto frontier analysis method. However, due to space constraints, the details of this method are not included in the present paper. Analysis revealed that, $Z = 2$ and $V = 512$, yielded the most favorable compromise between performance metrics and computational complexity overhead.

The subsequent sections present a comparative analysis of the performance metrics between our proposed data-driven approach, called deep multi-layer perceptron (DMLP)-based activity detector and a widely recognized maximum likelihood (ML) activity detector algorithm presented in \cite{ganesan2020clustering}. For performance evaluation, we employ the receiver operating characteristics (ROC) curve to assess the efficacy of the proposed approach. The ROC curve represents the variation of detection probability, $P_D$, with false alarm probability, $P_{FA}$.

Fig.~\ref{fig2} and Fig.~\ref{fig3} illustrate the performance of various device activity detection methods as function of system parameters $L$ and $\epsilon$, respectively. Fig.\ref{fig2} demonstrates, that for a fixed $P_{FA}$, the activity detection performance improves as the duration of pilot sequences increases. This result was expected due to the reduction in correlation between devices' pilot sequences as $L$ increases. Fig.\ref{fig3} shows that performance declines as activity levels $\epsilon$ ramp up, leading to higher cross-correlation between non-orthogonal pilot sequences. In both figures, the ML-based detector consistently outperforms the DMLP-based detector in scenarios with perfect channel estimation and inﬁnite ADC resolutions. This observed behavior may not extend to realistic scenarios with imperfect channel knowledge and finite ADC resolution, as we will illustrate in the next subsections.

\subsection{Resilience under Imperfect Channel Estimation}
In this investigation, we concentrate our analysis on the effects of channel estimation errors, which emerge as a predominant form of perturbation capable of substantially diminishing the performance of device activity detection algorithms. To accomplish this objective, we adopted the generic perturbation model described in \cite{marco2005validity} to dans  model the channel estimation errors. Therefore, the estimated large-scale fading coefficients 

\begin{equation}
\label{eq:Perturbation}
\tilde{\beta}_{mk} = \sqrt{1 - \theta^2} \beta_{mk} + \theta  \psi,
\end{equation}

where $\tilde{\beta}_{mk}$ denotes the perturbed version of the large-scale fading $\beta_{mk}$, and the noise term $\psi$, which follows a standard normal distribution. We assume that the noise is independent from $\beta_{mk}$. The parameter $\theta \in \left[0,1\right]$ indicates the quality of large-scale fading estimation ($i.e.$ perfect estimation is achieved when $\theta =0$).

\begin{figure}[t]
  \centering
  \includegraphics[width=\columnwidth, height=6cm]{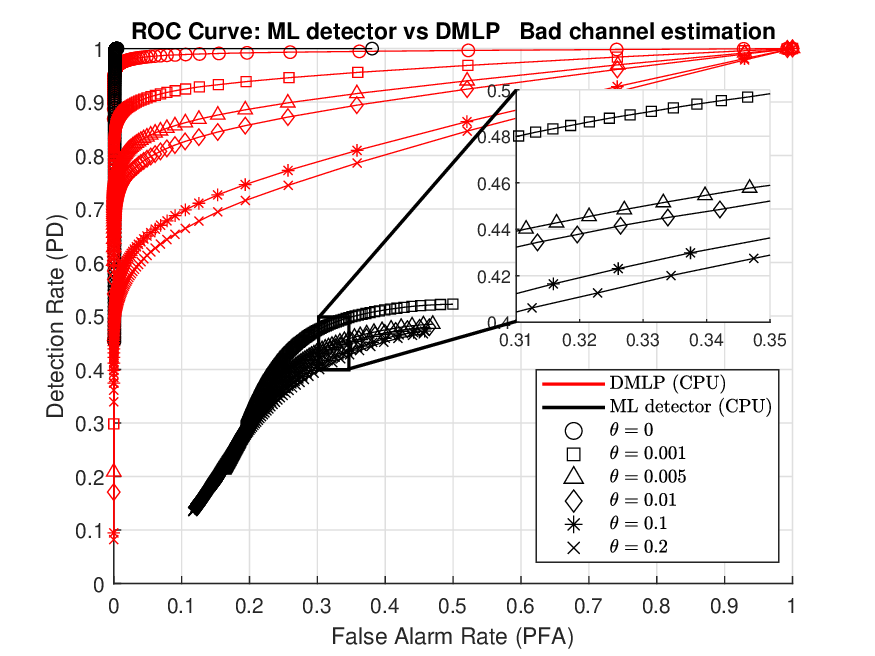}
  \vspace{-15pt} 
  \caption{ROC performance variation of both device activity detectors in presence of imperfect channel state information.}
  \label{Fig:BetaPert}
\end{figure}

The impact of imperfect channel estimation ($i.e.$ the estimated large-scale fading coefficient $\tilde{\beta}$ modeled as in Eq.\eqref{eq:Perturbation}) is given in Fig.~\ref{Fig:BetaPert}. We observe that the proposed DMLP-based approach yields the highest performance, whereas the performance of the ML-based activity detector is very poor. The DMLP utilizes $\beta_{mk}$ for the purpose of clustering, assigning each device to the most appropriate cluster. Using $\tilde{\beta}_{mk}$ coefficients can affect this clustering process performed by the CPU, potentially leading devices to be placed in a less optimal cluster. Despite the presence of channel estimation errors, it is evident that the DMLP-based activity detector continues to perform effectively, exhibiting only minor degradation in performance. Therefore, the DMLP-based approach exhibits significant stability despite the occurrence of errors in the selection process of the serving APs and cluster. In contrast, the performance of the ML-based detector declines significantly even in the presence of negligible estimation errors

The reliance on $\beta$ in the ML-based activity detector illustrates its sensitivity to performance decline when faced with channel estimation errors, a challenge that is less pronounced for the DMLP-based detector. This highlights the superior adaptability and robustness of the DMLP-baed detector in real-world scenarios, where perfect channel estimation may not always be accessible.

\begin{figure}[t]
  \centering
  \includegraphics[width=\columnwidth, height=6cm]{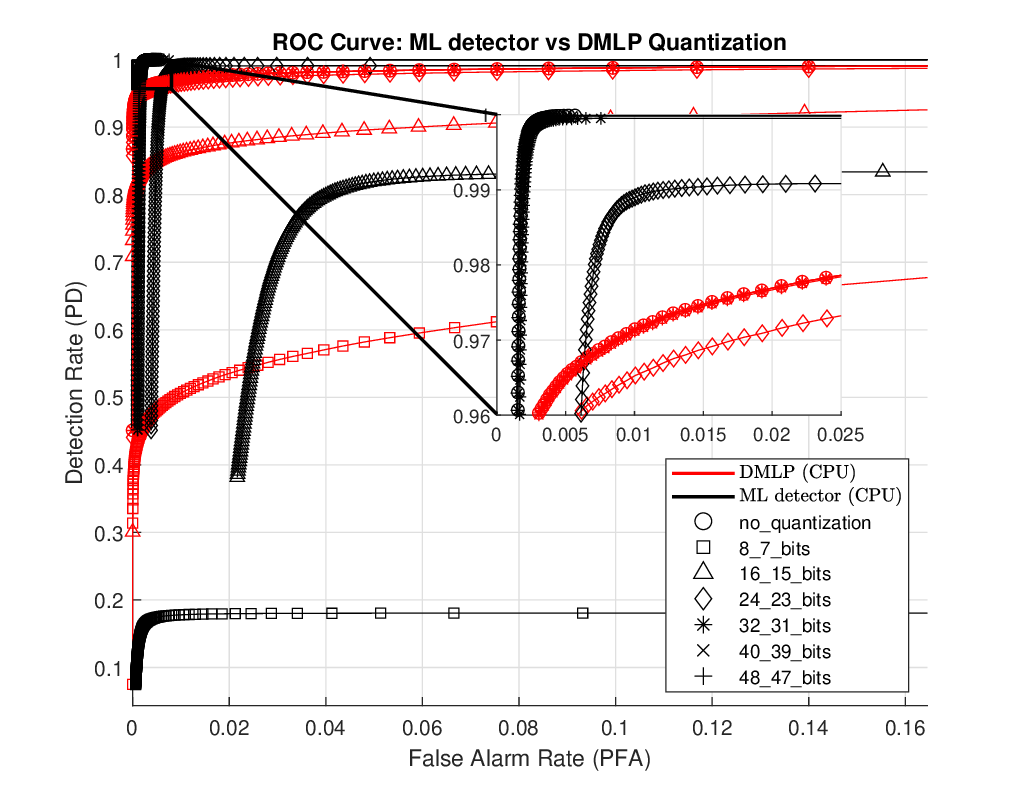}
  \vspace{-15pt} 
  \caption{ROC curves of both device activity detectors with fixed-point word lengths of large-scale fading $\beta_{mk}$, where $b\_q\_bits$ represents configuration where $b$ and $q$ denote the binary word length and the fractional-part size, respectively.}
  \label{Fig:Fixed_point_DMLP}
\end{figure}

\vspace{-5pt}
\protect\subsection{Resilience under finite ADC resolution} 
Assuming finite‐resolution analog-to-digital converters (ADCs) are employed at the access points. Herein, we investigate the effect of quantization noise, due to data-conversion process, on the performance metrics of the proposed DMLP-based activity detection approach. The process of converting from floating-point to fixed-point representation entails allocating a specified number of bits to digitally represent a given signal with infinite resolution. Within this framework, we evaluate how fixed-point conversion affects the performance of the DMLP-based and the ML-based detection methods. Before going into details, note that the ML-based method requires two inputs: received signals $\mathbf{Y}$ and perfect channels estimation. In contrast, the DMLP-based method only needs the received signals and can work with an imperfect channels estimation.

An extended simulation investigation was carried out on a variety of quantization configurations of the input received signals $\mathbf{Y}$, across both activity detection methods, demonstrates performance stability under quantization process.

Because the large-scale fading coefficient $\beta_{mk}$ values are widely dispersed ($i.e.$ the magnitude ranges from approximately $10^{-5}$ to $10^{-13}$), a substantial number of bits is necessary to achieve an accurate digital representation of this coefficient. In accordance with established fixed-point representation conventions, floating-point values will be expressed through a binary word of length $b=p+q$, where the first $p$ bits represent the integer-part while the last $q$ bits constitute the fractional-part. 

Fig.~\ref{Fig:Fixed_point_DMLP} reveals progressive performance enhancement of the ROC curve as the binary word length $b$ increases. This suggests that both detection methods performance benefits from a higher quantization resolution. Moreover, The curve indicates that performance enhancement is directly correlated with increased bit allocation to the fractional-part ($i.e.$ raise the value of $q$). When the precision is configured to $b=16$ bits, the performance of the DMLP is marginally lower than that observed with infinite-resolution, but it remains within acceptable limits. However, as the word-length $b$ diminishes, the DMLP-based method exhibits a marked performance deterioration—albeit less pronounced compared to the ML-based detector.

Fig.~\ref{Fig:Fixed_point_DMLP} illustrates a significant decline in performance for the ML-based detector as the fractional-part bits is reduced, emphasizing the critical importance of accurate $\beta_{mk}$ representation in fixed-point format. Additionally, the performance curves that most closely approximating the infinite resolution model are obtained at precision levels of $b=32$ bits or higher. 

Although the high precision level allows performance to closely approach floating-point accuracy, it also introduces several notable drawbacks. These include increased memory demands, higher computational resources, and greater power consumption. Such requirements can lead to a more complex system design, potentially affecting the system's overall efficiency especially in environments where resources are limited. While a high precision level brings performance closer to floating-point accuracy, it also introduces several notable challenges.

In summary, DMLP exhibits greater stability, as it is not directly affected by large-scale fading, resulting in only minimal degradation with reduced precision. In contrast, the ML-detector shows a clear dependence on whether the input undergoes fixed-point conversion, highlighting a significant difference in how device activity detection is performed.

\section{Conclusion}
\label{Section:Conclusion}
This paper investigates the performance of a machine learning-based approach for device activity detection in grant-free random access within cell-free massive MIMO networks. A data-driven algorithm is proposed that efficiently addresses the challenges of device activity detection, particularly in environments with imperfect channel state information and input quantization. The proposed method outperforms traditional techniques based on maximum likelihood in terms of accuracy and resilience to uncertainties, making it a promising candidate for future mMTC applications.
\section{Acknowledgements}
\noindent This work received funding from the French National Research Agency (ANR-22-CE25-0015) within the framework of the POSEIDON project.

\bibliographystyle{IEEEtran}
\bibliography{Main}

\begin{thebibliography}{10}
\providecommand{\url}[1]{#1}
\csname url@samestyle\endcsname
\providecommand{\newblock}{\relax}
\providecommand{\bibinfo}[2]{#2}
\providecommand{\BIBentrySTDinterwordspacing}{\spaceskip=0pt\relax}
\providecommand{\BIBentryALTinterwordstretchfactor}{4}
\providecommand{\BIBentryALTinterwordspacing}{\spaceskip=\fontdimen2\font plus
\BIBentryALTinterwordstretchfactor\fontdimen3\font minus \fontdimen4\font\relax}
\providecommand{\BIBforeignlanguage}[2]{{%
\expandafter\ifx\csname l@#1\endcsname\relax
\typeout{** WARNING: IEEEtran.bst: No hyphenation pattern has been}%
\typeout{** loaded for the language `#1'. Using the pattern for}%
\typeout{** the default language instead.}%
\else
\language=\csname l@#1\endcsname
\fi
#2}}
\providecommand{\BIBdecl}{\relax}
\BIBdecl

\bibitem{Shahab2020}
M.~B. Shahab, R.~Abbas, M.~Shirvanimoghaddam, and S.~J. Johnson, ``Grant-free non-orthogonal multiple access for iot: A survey,'' \emph{IEEE Communications Surveys \& Tutorials}, vol.~22, no.~3, pp. 1805--1838, 2020.

\bibitem{bai2018deep}
Y.~Bai, B.~Ai, and W.~Chen, ``Deep learning based fast multiuser detection for massive machine-type communication,'' in \emph{2019 IEEE 90th VTC2019-Fall}, 2019, pp. 1--5.

\bibitem{deSouza2023DeepLearning}
J.~H.~I. de~Souza and T.~Abrão, ``Deep learning-based activity detection for grant-free random access,'' \emph{IEEE Systems Journal}, vol.~17, no.~1, pp. 940--951, 2023.

\bibitem{Ganesan2020Algorithm}
U.~K. Ganesan, E.~Bjornson, and E.~G. Larsson, ``An algorithm for grant-free random access in cell-free massive mimo,'' in \emph{2020 IEEE 21st International Workshop on SPAWC}, 2020, pp. 1--5.

\bibitem{ganesan2020clustering}
U.~K. Ganesan, E.~Björnson, and E.~G. Larsson, ``Clustering-based activity detection algorithms for grant-free random access in cell-free massive mimo,'' \emph{IEEE Transactions on Communications}, vol.~69, no.~11, pp. 7520--7530, 2021.

\bibitem{Cell-free-lowADC}
X.~Hu, C.~Zhong, X.~Chen, W.~Xu, H.~Lin, and Z.~Zhang, ``Cell-free massive mimo systems with low resolution adcs,'' \emph{IEEE Transactions on Communications}, vol.~67, no.~10, pp. 6844--6857, 2019.

\bibitem{Beta1}
C.~Wang, O.~Y. Bursalioglu, H.~Papadopoulos, and G.~Caire, ``On-the-fly large-scale channel-gain estimation for massive antenna-array base stations,'' in \emph{2018 IEEE International Conference on Communications (ICC)}, 2018, pp. 1--6.

\bibitem{Beta2}
A.~Fengler, S.~Haghighatshoar, P.~Jung, and G.~Caire, ``Non-bayesian activity detection, large-scale fading coefficient estimation, and unsourced random access with a massive mimo receiver,'' \emph{IEEE Transactions on Information Theory}, vol.~67, no.~5, pp. 2925--2951, 2021.

\bibitem{Bjornson2019}
E.~Björnson and L.~Sanguinetti, ``Making cell-free massive mimo competitive with mmse processing and centralized implementation,'' \emph{IEEE Transactions on Wireless Communications}, vol.~19, no.~1, pp. 77--90, 2020.

\bibitem{marco2005validity}
D.~Marco and D.~L. Neuhoff, ``The validity of the additive noise model for uniform scalar quantizers,'' \emph{IEEE Transactions on Information Theory}, vol.~51, no.~5, pp. 1739--1755, 2005.

\end{thebibliography}

\end{document}